\documentclass[aps,prb,groupaddress,twocolumn,floatfix]{revtex4-2}
\usepackage{units}
\usepackage{amsmath}
\usepackage{amssymb}
\usepackage{graphicx}
\usepackage{bm}
\usepackage{multirow,color,relsize,ulem,microtype}

\newcommand{\be}{\begin{equation}}
\newcommand{\ee}{\end{equation}}

\newcommand{\re}[1]{\text{Re}[#1]}
\newcommand{\im}[1]{\text{Im}[#1]}

\begin{document}

\title{Topological chiral edge modes in the continuum of a trivial bulk}

\author{Li Ge}
\affiliation{Department of Physics and Astronomy, College of Staten Island, CUNY, Staten Island, NY, USA}
\affiliation{The Graduate Center, CUNY, New York, NY, USA}

\begin{abstract}
Chiral edge modes in topological states of matter are routinely found inside band gaps, which may be deformed and shifted as a function of the lattice momentum. Here we show that they can exist inside a bulk band instead, i.e., in the continuum. These chiral edge modes in the continuum (CEMICs) are induced by introducing an imaginary magnetic flux to an otherwise topologically trivial two-dimensional system, realized by imposing asymmetric couplings along the edge. While the stronger couplings provide a preferred circulation direction, we find that the edge localization of CEMICs does \textit{not} originate from the non-Hermitian skin effect, which is typically associated with systems featuring asymmetric couplings and the open boundary condition. Instead, this edge localization is due to the energy exchange between the system and the environment at the asymmetric coupling junctions, which leads to a parity-time (PT) transition in the Brillouin zone. Intriguingly, the onset of CEMICs is independent of the system size in the macroscopic limit, which is given by the golden ratio between the asymmetric couplings and their geometric average. 
\end{abstract}

\maketitle

\section{introduction}

Chiral edge modes are unidirectional and boundary-localized excitations that arise in systems with nontrivial topological orders \cite{hasan_colloquium_2010,qi2011topological,Alicea,Beenakker,sarma_majorana_2015,Alicea2}. For example, chiral edge modes in a Chern insulator exist in a topological band gap, where the difference between the Chern numbers above and below the band gap determines the number of edge bands. Similar chiral edge modes also exist in band gaps of the valley Hall insulator due to opposite valley Chern numbers \cite{Xiao2007,peng_topological_2024}. While such a band gap may be deformed and shifted \cite{colomes_antichiral_2018} as a function of the lattice momentum and are hence ``incomplete'' across the entire Brillouin zone \cite{Yablonovitch1987,John1987}, a finite energy separation is still required between the edge and bulk bands at the carrier momentum of a chiral wave packet in order for it to circulate along the edge without leaking into the bulk, such as in Chern metals and semimetals \cite{zhou_observation_2024,chen_chern-protected_2025}. 
This reliance of chiral edge modes on a band gap was observed not only in Hermitian systems (such as condensed matters and cold atoms \cite{Goldman2016}) but also in non-Hermitian systems (such as photonics \cite{lu2014topological,bandres_topological_2018,zhao2019non,rivero_pseudochirality_2020} and acoustics \cite{xue_topological_2022}). 

In the meanwhile, a wide range of interesting physics also take place inside bulk bands (i.e., the continuum), and a particular phenomenon relevant for our discussion below is known as bound states in the continuum (BICs) \cite{hsu_bound_2016}, which has attracted fast-growing attention in photonics due to its application for single-mode lasers with an ultra-high quality factor \cite{huang_ultrafast_2020}. In comparison, the continuum where a photonic BIC exists has a finite and typically low quality factor due to the radiation loss of a photonic crystal. 

In this work, we show that chiral edge modes can also exist inside a trivial bulk band, or in other words, in the continuum of a trivial bulk. More specifically, its frequency is not gapped from the bulk states, independent of the lattice momentum; the gap, instead, exists in the decay rate or imaginary part of the energy, similar to a BIC. While such an imaginary gap has been used to explain zero-dimensional topological states \cite{gong_topological_2018} and introduce new boundaries in two-dimensional topological systems \cite{zhao2019non,SciPostPhys21}, it has not been attributed as the source of chiral edge modes in an otherwise topologically trivial system.  

We exemplify these chiral edge modes in the continuum (CEMICs) using a tight-binding square lattice, in a ribbon geometry as well as a finite square domain. In both cases, the couplings are symmetric in the bulk but asymmetric along the edges, leading to an imaginary magnetic flux through each boundary plaquette and the whole system. We demonstrate the robustness of CEMICs through several examples, including the suppression of back- and bulk-scattering caused by trivial defects, enabled by the imaginary band gap, and the observation of bulk-assisted traversing of topological defects, enabled by the frequencies of these chiral edge modes that are in the continuum.   

Unlike the previously studied hybrid skin-topological effect \cite{sun_photonic_2024,jiang_observation_2024,liu_localization_2024}, our system is topologically trivial and has no chiral edge states before the introduction of these asymmetric couplings. More importantly, the emergence of CEMICs is unrelated to the non-Hermitian skin effect \cite{hatano_localization_1996,hatano_vortex_1997}, which is typically associated with systems featuring asymmetric couplings and the open boundary condition. Instead, our CEMICs are due to the energy exchange between the system and the environment at the asymmetric coupling junctions, which leads to a parity-time (PT) transition in the Brillouin zone \cite{feng2017non,el2018non,ashida2020non,ozdemir_paritytime_2019}. Intriguingly, the onset of CEMICs is independent of the system size in the macroscopic limit, which is given by the golden ratio between the asymmetric couplings and their geometric average.

\section{Models and Results}

\begin{figure}[t]
\includegraphics[clip,width=\linewidth]{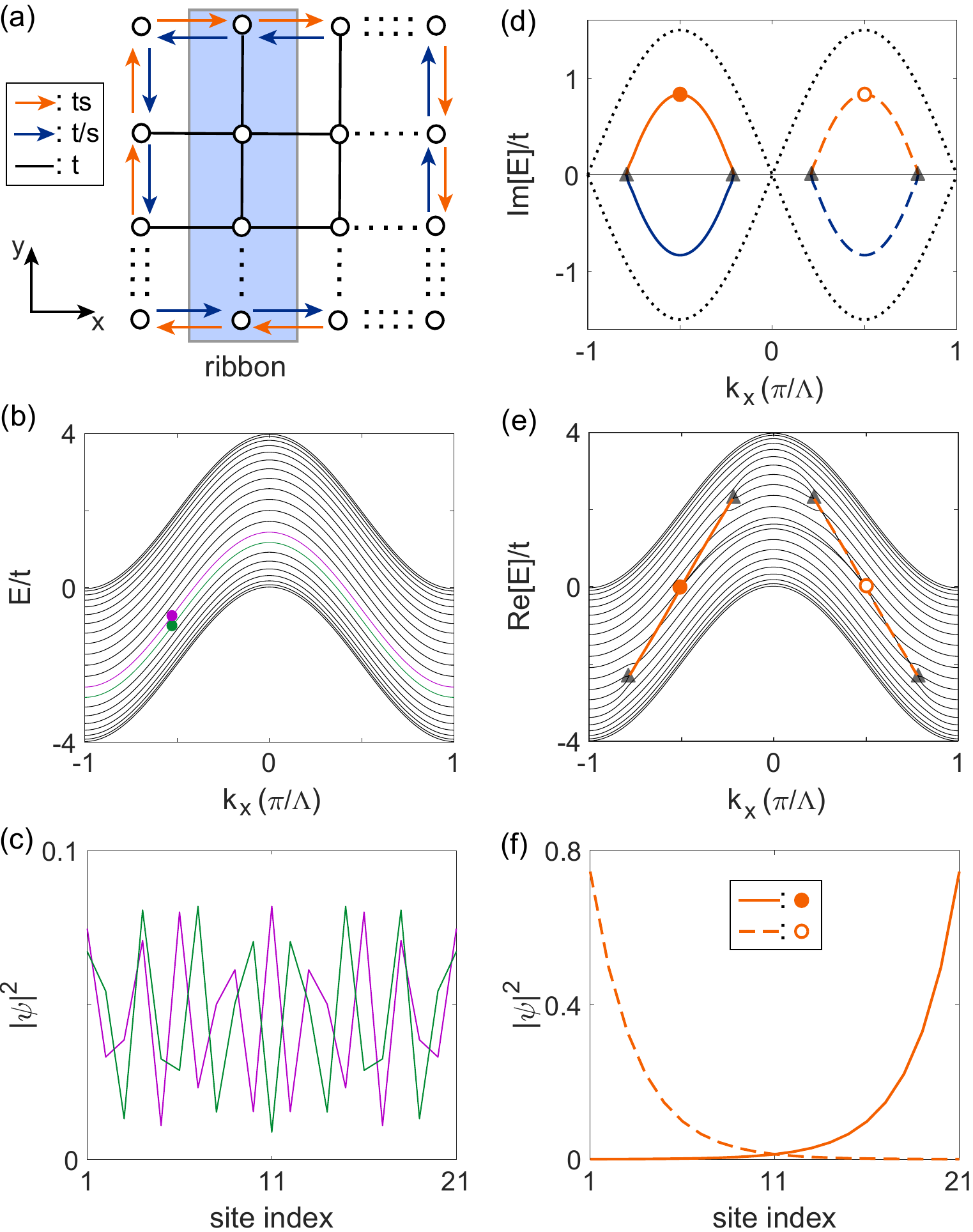}
\caption{\textbf{Onset of CEMICs in a ribbon model}. (a) Schematics showing the square lattice in a square domain and a vertical ribbon. Dotted lines indicate the omitted part. (b) Band structure of the vertical ribbon in (a) with 21 sites and $s=1.2$. The wave functions of bands 8 and 9 from the bottom at $k_x=-\pi/2\Lambda$ (dots) are shown in (c). (d,e) Imaginary and real parts of the band structure with $s=2$. The wave functions of the two CEMICs marked by the two large dots are shown in (f). Dotted lines in (d) show the upper limit of the effective gain and loss given by Eq.~(\ref{eq:ImE}). Triangles in (d) and (e) indicate EPs.} \label{fig:1}
\end{figure}

The tight-binding square lattice mentioned above is illustrated in Fig.~\ref{fig:1}(a), where $\Lambda$ is the lattice constant in both directions. The asymmetric nearest-neighbor couplings along the edges are given by $ts$ and $t/s$ in the clockwise (CW) and counterclockwise (CCW) directions, respectively. $t>0$ is their geometric average, and it equals the symmetric coupling in the bulk. $s$ is positive number, which indicates a stronger CW (CCW) hopping when $s>1$ ($s<1$). An effective imaginary magnetic flux threads through every domain where such CW or CCW asymmetric couplings are formed \cite{hatano_vortex_1997}, including each boundary plaquette and the whole system.

\subsection{CEMICs}
\label{sec:cemics}

Naively, one would expect a chiral edge mode in the direction of the stronger coupling, which, however, is not the case when $s$ is close to 1. We exemplify the lack of chiral edge modes in Fig.~\ref{fig:1}(b) at $s=1.2$ using a ribbon, i.e., a single column in the middle of the square domain [see Fig.~\ref{fig:1}(a)], with the periodic boundary condition (PBC) in the horizontal direction. Here we consider a ribbon with $N_y=21$ sites in the vertical direction, which gives a single bulk band with 21 modes at a given $k_y$. They are smooth across the entire Brillouin zone and do not display degeneracies or singularities. Correspondingly, their wave functions are extended in the vertical direction [see the examples in Fig.~\ref{fig:1}(c) at $k_x=-\pi/2\Lambda$] and do not exhibit edge localization. 

As we further increase $s$, however, CEMICs emerge as we show in Fig.~\ref{fig:1}(e) (colored lines), which are completely embedded in the bulk band. Their wave functions are exemplified in Fig.~\ref{fig:1}(f): The right-moving CEMIC [i.e., with a positive slope in Fig.~\ref{fig:1}(e)] at $k_x=-\pi/2\Lambda$ is localized at the top edge, while the left-moving CEMIC at $k_x=\pi/2\Lambda$ is localized at the bottom edge. To avoid confusion, let us emphasize that the group velocity of these two CEMICs has the \textit{opposite} sign of the lattice momentum.

This finding of a chiral edge mode in the direction of stronger coupling is consistent with the naive expectation mentioned previously. However, we also find two CEMICs moving in the direction of \textit{weaker} coupling, i.e., with the right-moving (left-moving) one localized at the bottom (top) edge. In fact, each colored line shown in Fig.~\ref{fig:1}(e) represents two CEMICs that share the same group velocity (i.e., they share the same $\re{E(k_x)}$) but are localized on opposite edges. They are distinguished by the imaginary part of the energy eigenvalue: They are amplified (attenuated) as they propagate in the direction of strong (weaker) couplings. For example, the two previously mentioned wave functions shown in Fig.~\ref{fig:1}(f) are for the two CW CEMICs with the maximum gain, marked by the two large dots in Fig.~\ref{fig:1}(d). The other two CEMIC bands with loss are in the CCW direction, with their loss shown by the solid and dashed blue lines in Fig.~\ref{fig:1}(d). 

\subsection{Origin of gain and loss}
\label{sec:gain_and_loss}

We note that our ribbon model does not have explicit gain or loss, i.e., an imaginary on-site potential. The origin of the gain and loss manifested by the CEMICs, therefore, is the implicit energy exchange with the environment at the asymmetric coupling junctions \cite{ge_non-hermitian_2023,lyu_non-hermitian_2024}. To highlight this behavior, we first assume that the top and bottom edges are isolated from the bulk, i.e., they are two independent one-dimensional (1D) Hatano-Nelson chains. These tight-binding chains along each edge lead to the following equation for the probability amplitude:
\be
\frac{d|\psi_n|^2}{dt} = {\cal J}_{n,n+1} + {\cal J}_{n,n-1}.\label{eq:dynamics}
\ee
Here $n$ is the column index increasing from left to right, and 
\be
{\cal J}_{n,n\pm1} = it_{n,n\pm1}^*\psi^*_{n\pm1} \psi_{n} + c.c.\nonumber
\ee
are the inter-site probability currents from sites $n\pm1$ to $n$. If the probability current is conserved, then the particles leaving site $n$ and going to the left (right) should all enter site $n-1$ ($n+1$). As a result, we find that  
${\cal J}_{n,n-1} =- {\cal J}_{n-1,n}$ (${\cal J}_{n,n+1} = - {\cal J}_{n+1,n}$). If these identities are broken, then there must be particle or energy exchange with the environment at the \textit{coupling junctions}.

Therefore, we define the energy gained from the environment at the coupling junction between sites $n$ and $n+1$ by
\be
g_{(n,n+1)} \equiv {\cal J}_{n,n+1} + {\cal J}_{n+1,n}, 
\ee
and we note that this quantity can vanish even with asymmetric couplings, which is the case in the 1D Hatano-Nelson model with the open boundary condition (OBC): With real asymmetric couplings $t_{n,n+1}\neq t_{n+1,n}$, the wave functions in the 1D Hatano-Nelson model with OBC are all real-valued, and hence ${\cal J}_{n,n\pm1}$ and $g_{(n,n+1)}$ vanish. The lack of energy exchange with the environment is in fact one way to understand the real energy spectrum of this model, with the other being the imaginary gauge transformation \cite{hatano_localization_1996}.

With PBC along the horizontal direction, i.e., $\psi_{n\pm1} = \psi_n e^{\pm ik_x \Lambda}$, we force the wave function to be complex-valued, and we
immediately find that ${\cal J}_{n,n\pm1} = \pm2t_{n,n\pm1}\sin(k_x\Lambda)|\psi_n|^2$, leading to  
\be
g_{n,n+1} = 2(t_{n,n+1} - t_{n+1,n})\sin{k_x\Lambda}|\psi_n|^2.
\ee
Along the top edge we have $t_{n,n+1}=t/s,\, t_{n+1,n} = ts$, and with $s>1$, we find $g_{(n,n+1)}$ to be positive (negative) when $k_x<0$ ($k_x>0$), meaning that the environment injects energy into the right-moving CEMIC [solid orange line in Fig.~\ref{fig:1}(d)] and extracts energy from the left-moving CEMIC [dashed blue line in Fig.~\ref{fig:1}(d)]. The energy exchange with the environment is reversed at the bottom edge because the couplings are given by $t_{n,n+1}=ts,\, t_{n+1,n} = t/s$ instead. 

Now with the internal sites of the ribbon (and the square domain) also considered, the upper and lower edges share the effective gain or loss with the rest of the lattice. The upper bound of the effective gain and loss coefficients in the CEMIC bands [dotted lines in Fig.~\ref{fig:1}(d)] can then be derived from Eq.~(\ref{eq:dynamics}), i.e.,  
\be
\im{E_b} =  \frac{g_{(n,n+1)}}{2|\psi_n|^2} = \mp \left(s - \frac{1}{s}\right)t\sin{k_x\Lambda}\equiv\pm \gamma \label{eq:ImE}
\ee 
using $\psi_n\propto e^{-iE\tau}\,(\hbar=1)$.

\subsection{Onset of CEMICs}

This picture of energy exchange with the environment, however, does not tell the whole story: It does not explain the onset of the CEMIC bands above a certain asymmetric coupling factor $s$, as we have seen in Fig.~\ref{fig:1}. We denote this critical value by $s_c$, and we find it to be close to 1.62 numerically in the macroscopic limit (i.e., with $N_y\gg1$). Below we show that this value is given by the golden ratio $(1+\sqrt{5})/2$ and corresponds to a PT transition \cite{bender_real_1998,feng2017non}. 

Let us first study a qualitative picture using the two right-moving CEMICs, one along the top edge and the other along the bottom edge. Not only do they couple to the bulk, but they also couple to each other indirectly via the bulk. Hence we can formally write down the classical $2\times2$ PT Hamiltonian for these two CEMICs by eliminating the bulk:
\be
H_e(k_x) = \begin{pmatrix}
\re{E} + i\gamma & t_e \\
t_e & \re{E} - i\gamma
\end{pmatrix}.\label{eq:He}
\ee
The \textit{effective} gain and loss $\gamma$ given by Eq.~(\ref{eq:ImE}) now appear \textit{explicitly} as an imaginary on-site potential at the upper and lower edge of the ribbon, and we have dropped the $k_x$-dependence of all quantities on the right, including $\gamma$ and the effective coupling $t_e$ between the two CEMICs.  

With a small $s$, the maximum of $\gamma$ (i.e., at $k_x=\pm\pi/2\Lambda$) is weaker than $t_e$, and hence the reduced $2\times 2$ system stays in the PT-symmetric phase, which gives two bands with real energies and wave functions with equal probability amplitudes along the two edges [cf. Fig.~\ref{fig:1}(c), for example]. With a larger $s$, the value of $\gamma$ close to $k_x=\pm\pi/2\Lambda$ becomes larger than $t_e$, and they lead to a PT-broken phase, with complex energies and wave functions localized at the two edges [cf. Fig.~\ref{fig:1}(f), for example]. Close to $k_x=0$ and $\pm\pi/\Lambda$, the $\sin(k_x\Lambda)$-dependence of $\gamma$ determines that the coupling still dominates, and the reduced system is still in the PT-symmetric phase, which are separated from the PT-broken phases by exceptional points (EPs) [see the triangles in Figs.~\ref{fig:1}(d) and \ref{fig:1}(e)]. 
As we further increase $s$, the PT-broken phases broaden and push the EPs towards $k_x=0$ and $\pm\pi/\Lambda$ (see Appendix~\ref{app:couplings}).

To be more quantitative and reveal the relationship between $s_c$ and the golden ratio, we revisit the ribbon model, which is a tridiagonal matrix with $t$'s along the first upper and lower diagonals. On the main diagonal, the elements are $2t\cos k\Lambda$ except for the first and last elements, given by  $(s+1/s)t\cos k\Lambda\pm i\gamma$, where $\gamma$ is defined in Eq.~(\ref{eq:ImE}). Next, we focus on $k_x=\mp\pi/2\Lambda$ where the effective gain and loss given by $\gamma$ are the strongest (and hence also where the PT transition is most likely to take place at a small $s$). It is easy to see that the ribbon Hamiltonian now becomes simply
\be
H_c = t\begin{pmatrix}
+i(s-1/s) & 1 & & & \\
         1 & 0 & 1 & & \\
           & \ddots & \ddots & \ddots & \\
           & & 1 & 0 & 1 \\
           & &  & 1  & -i(s-1/s)
\end{pmatrix}.\label{eq:Hc}
\ee 
for the right-moving waves (i.e., $k=-\pi/2\Lambda$). For the left-moving waves, $H_c$ is given by switching the effective gain and loss, i.e., flipping the signs in the first and last diagonal elements.

The EP of $H_c$ can be shown to be at $s-1/s=1$ when $N_y$ is even, independent of the magnitude of $N_y$. In the case that $N_y$ is odd, the condition that leads to an EP converges to the same expression in the macroscopic limit (i.e., $N_y\gg 1$; see Appendix~\ref{app:EP}). In other words, the onset of CEMICs is independent of the system size in the microscopic limit, and the critical value $s_c>0$ obtained from this relation above is exactly the golden ratio $(1+\sqrt{5})/2$. 

We also note that the coalesced wave function at this EP is always uniform in space when $N_y$ is even and approaches uniform when an odd $N_y$ becomes much greater than 1 (see Appendix \ref{app:EP}). Although the PT-symmetric phase of $H_c$ does not rule out the possibility of an eigenstate localized at both the top and bottom edges, the latter does not occur in $H_c$ above, for either right- or left-moving waves. Therefore, CEMICs in each $H_c$ can only emerge in the PT-broken phase, with $s>s_c$ and localized at either the top or bottom edge.  


\subsection{A square domain}

Having demonstrated CEMICs in the ribbon model, we now show their existence in the square domain, with a finite $N_x=N_y\equiv N$. 

\begin{figure}[t]
\includegraphics[clip,width=\linewidth]{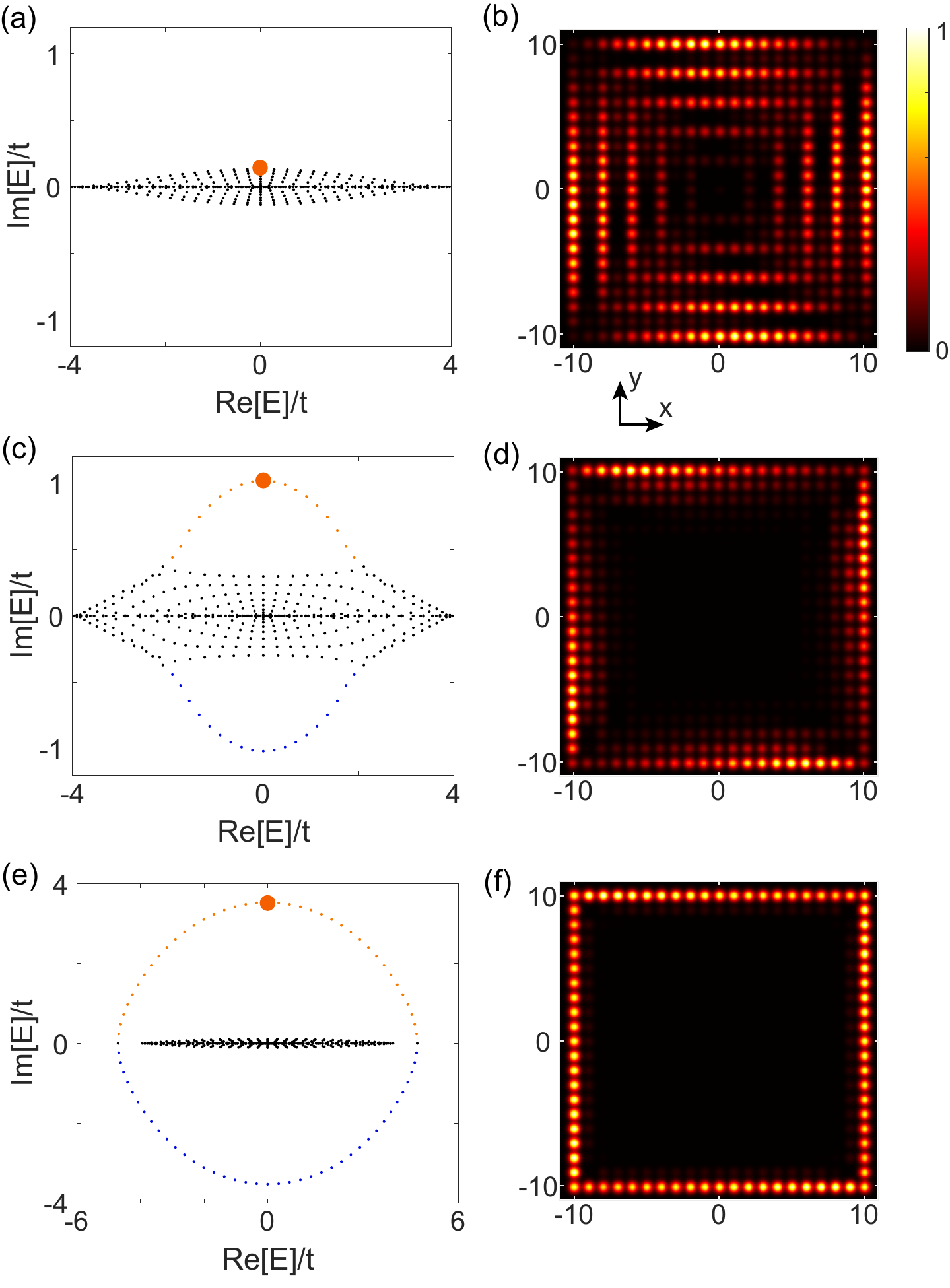}
\caption{\textbf{Onset of CEMICs in a square domain}. (a,c,e) The complex energy spectra at $s=1.2$, $2$, and $4$. There are $21$ sites in both the $x$ and $y$ directions. (b,d,f) Modes with the highest gain in these three cases [filled circles in (a,c,e)], which are bulk-like and CEMICs respectively.} \label{fig:square}
\end{figure}


Due to the degeneracy of the square lattice in a square domain, the spectrum of the latter becomes complex as soon as $s$ is greater than 1 [Fig.~\ref{fig:square}(a)]: The system has at least two Hermitian zero modes with $E=0$, which undergo a thresholdless PT transition \cite{ge_parity-time_2014} as $s$ becomes larger than 1. One can easily check the validity of this reasoning in the small system limit, e.g., with just two sites in the $x$ and $y$ directions respectively. Now the bulk in the square domain is absent, and we end up with a Hatano-Nelson model with 4 sites and the periodic boundary condition. As soon as $s$ becomes greater than 1, the two Hermitian zero modes (occupying the diagonal and anti-diagonal respectively) become non-Hermitian, with their energies given by $E=\pm it(s-1/s)$ on the imaginary axis.

Despite its complex spectrum, the modes of the square domain remain bulk-like until $s$ becomes comparable or much larger than $s_c$ [cf. Figs.~\ref{fig:square}(b) and (d)], where the imaginary part of the edge bands again becomes separated from the bulk bands [Fig.~\ref{fig:square}(c)]. We note that the topology of the CEMICs is inherited from the Hatano-Nelson model: As $s$ further increases, the edge band detaches from the bulk band. The former is given approximately by the 1D Hatano-Nelson model and forms an ellipse in the complex plane, i.e., $\re{E}^2/(s+1/s)^2 + \im{E}^2/(s-1/s)^2=t^2$ [Fig.~\ref{fig:square}(e)], while the bulk band is restricted to near the real axis with $|\re{E}|<4t$. As a result, the edge sites are also effectively detached from the bulk lattice, resulting in an almost uniform probability distribution in any of the edge modes [Fig.~\ref{fig:square}(f)].

\begin{figure}[b]
\includegraphics[clip,width=\linewidth]{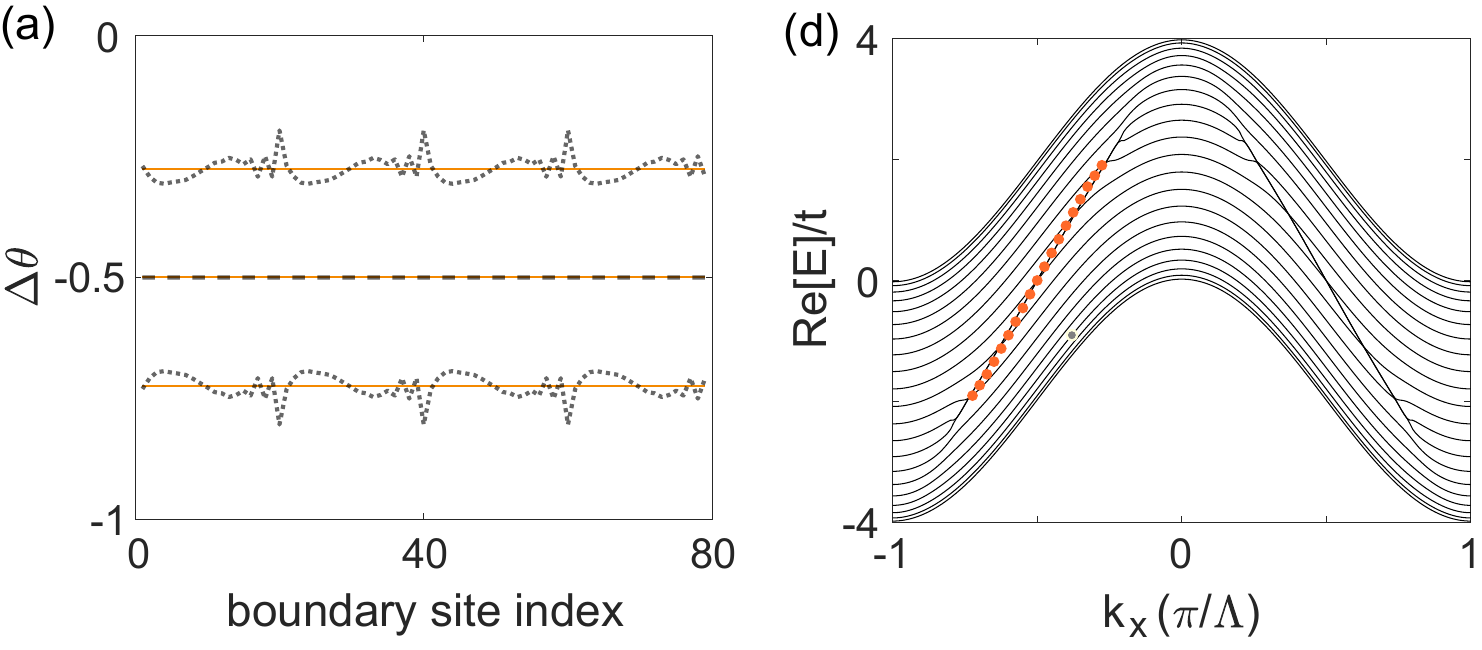}
\caption{\textbf{Characterizing CEMICs in the square domain}. (a) Consecutive phase difference in three CEMICs along the edge of the square domain. (b) Dispersion relation $\re{E(\bar{k})}$ from (a) and superposed with the band structure from the ribbon model.} \label{fig:phase}
\end{figure}

To show that the CEMICs obtained in the square domain agree well with the ribbon model, we calculate the phase difference between two consecutive sites along the perimeter of the square domain in the CEMIC band with gain [orange dotted line in Fig.~\ref{fig:square}(c)]. This quantity is denoted by $\Delta\theta$ in Fig.~\ref{fig:phase}(a), and it oscillates around a mean value $\bar{k}\Lambda$ for each mode inside this CEMIC band. We then use $\bar{k}$ and $E$ of each CEMIC to plot the dispersion relation $\re{E(\bar{k})}$, which agrees well with the band structure calculated from the ribbon model [see Fig.~\ref{fig:phase}(b)]. The same holds for the CEMICs with loss (not shown). 

\subsection{Dynamics and robustness of CEMICs}

\begin{figure}[t]
\includegraphics[clip,width=\linewidth]{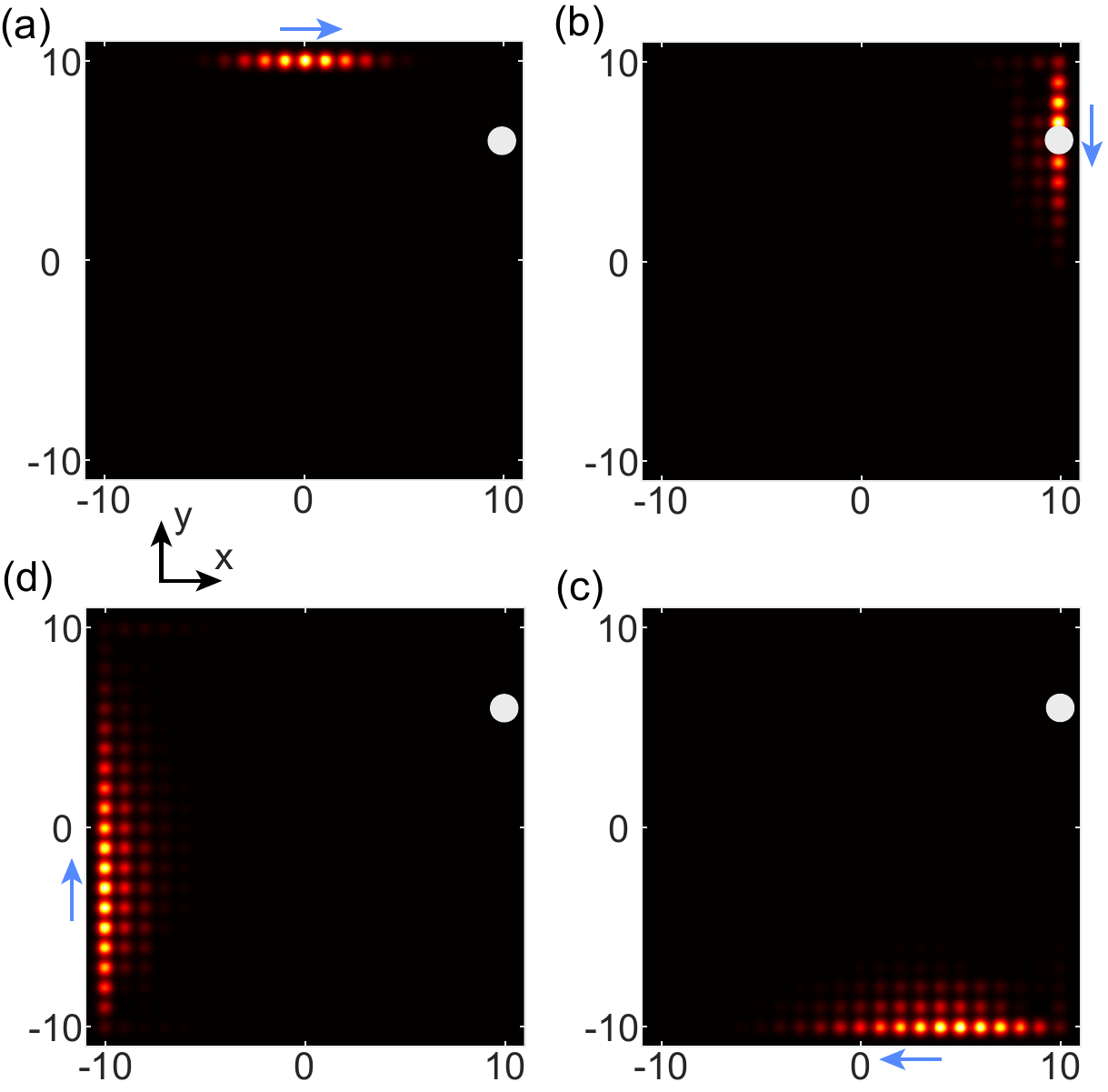}
\caption{\textbf{Propagation of the CW CEMIC with a defect at the top edge}. The snapshots are taken at time $\tau=0, 4.4, 12.4, 20.4$ in units of the inverse coupling $t^{-1}$, each normalized by $\text{max}|\psi_n|^2=1$. The width of the initial packet is $\sigma=4\Lambda$. The other parameters and color scheme are the same as in Figs.~\ref{fig:square}(c,d).} \label{fig:square_dynamics}
\end{figure}

With $s>1$, the CEMIC with the highest gain has a lattice momentum of $k_x=-\pi/2\Lambda$. 
We hence construct an initial wave packet $\psi_0(x) = e^{-x^2/\sigma^2-i\pi x/2\Lambda}$ along the top edge (Fig.~\ref{fig:square_dynamics}), and the dynamics shows a robust CEMIC traveling in the anticipated CW direction, despite the presence of the defect at the right edge with the onsite potential set to $t$. It should be noted that this defect does scatter the CW CEMIC into the bulk and the opposite CEMIC, but the weaker gain and even loss of these waves suppress their presence.

We also note that the group velocity given by the slope of the CEMIC band is almost a constant [see, for example, Fig.~\ref{fig:phase}(c)]. If we approximate it using $v_g=dE/dk = (s+1/s)\Lambda t$ obtained in the large $s$ case at $k_x=-\pi/2\Lambda$ (see Appendix~\ref{app:couplings}), the traverse time along one side, i.e., $(N-1)\Lambda/v_g$, is roughly $8$ in the units of inverse coupling $t^{-1}$ for the case shown in Fig.~\ref{fig:square_dynamics}. This behavior is qualitatively captured in the simulation, despite the broadening of the wave packet due to the group dispersion (i.e., $d^2 E/dk^2\neq 0$) and the presence of the corners. 

\begin{figure}[t]
\includegraphics[clip,width=\linewidth]{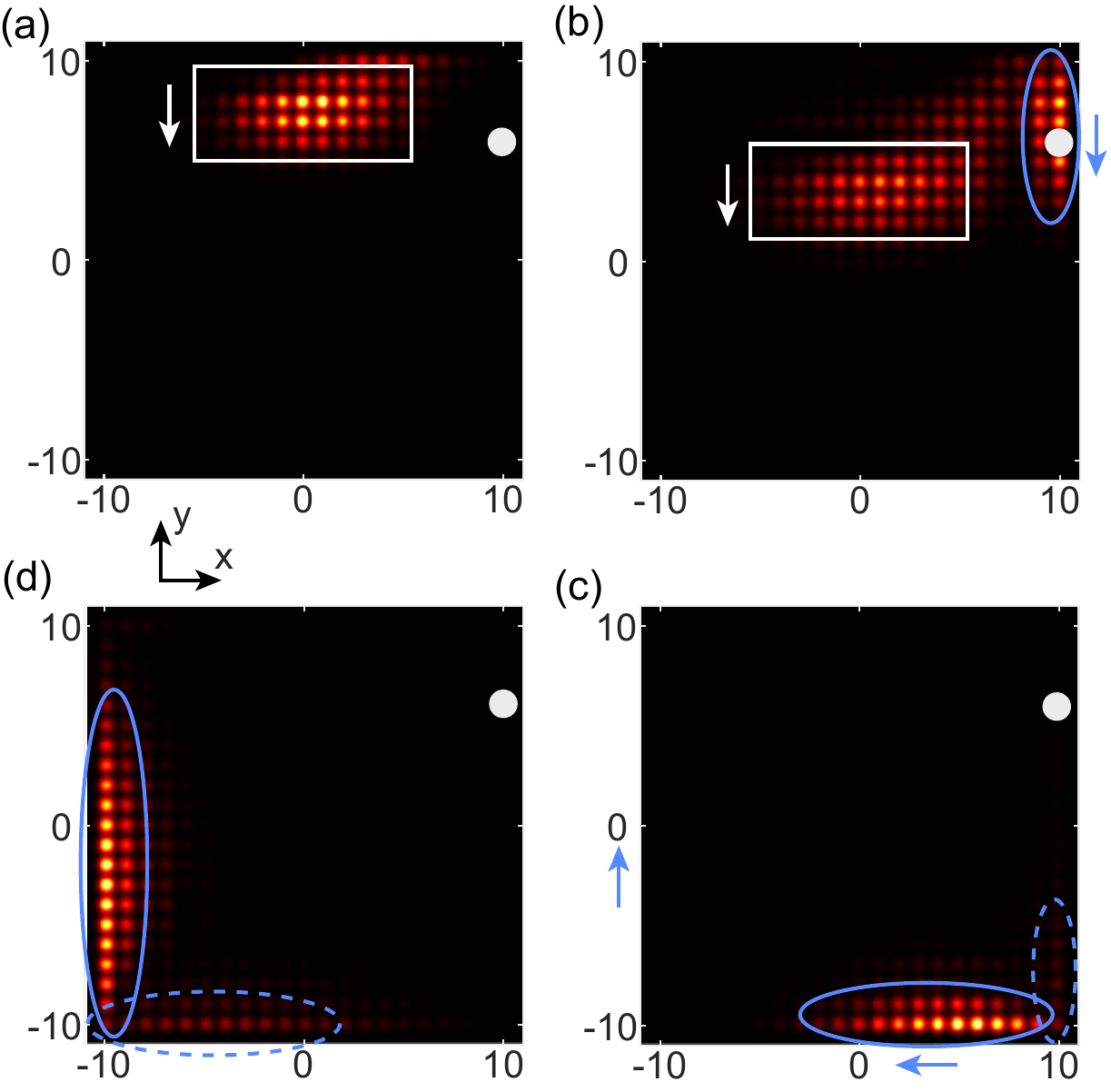}
\caption{\textbf{Dynamics of the CW CEMIC with bulk excitation}. All panels are the same as Fig.~\ref{fig:square_dynamics} except for the initial pulse momentum $k_x=0$ and the first frame at $\tau=2.2 t^{-1}$. Box, solid ellipse, dashed ellipse indicate bulk wave, CEMIC and its tail, respectively.} \label{fig:square_dynamics_bulk}
\end{figure}

To further show the robustness of CEMIC and its existence in the bulk continuum, we start with another wave packet with the same initial intensity profile but centered at $k_x=0$, where the CEMICs do not exist at $s=2$. As a result, we observe the propagation of the initial wave packet into the bulk along the $-y$ direction [boxed in Fig.~\ref{fig:square_dynamics_bulk}(a)]. However, the initial spread of the wave packet about $k_x=0$ causes an overlap with both the CW and CCW CEMIC bands. Due to its high gain, the former is amplified to a comparable intensity to the bulk wave after the first corner [see Fig.~\ref{fig:square_dynamics_bulk}(b)]. This edge component becomes dominant as the wave further propagates [Fig.~\ref{fig:square_dynamics_bulk}(c)], and it becomes almost identical to that in Fig.~\ref{fig:square_dynamics}(c), except for the trailing edge that becomes more noticeable in Fig.~\ref{fig:square_dynamics_bulk}(d).

\begin{figure}[t]
\includegraphics[clip,width=\linewidth]{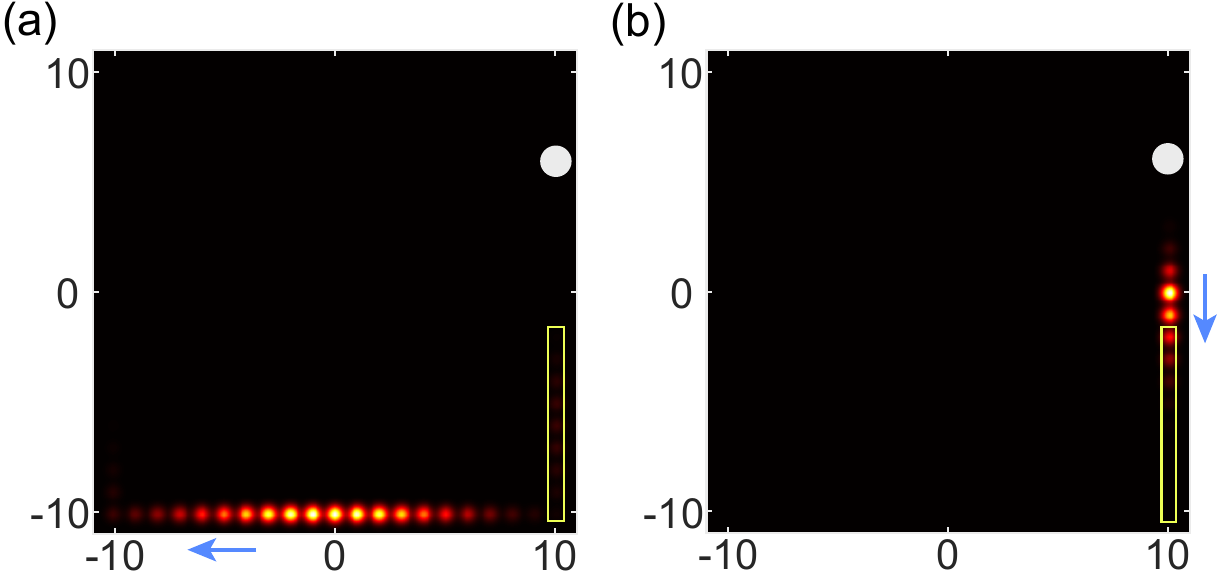}
\caption{\textbf{Propagation of a pulse in an edge chain}. The initial pulse and point defect are the same as in Fig.~\ref{fig:square_dynamics}, with the addition of the topological defect with symmetric coupling $t$ in the yellow ribbon. The snapshots are taken at time $\tau=12.4$ and $16.4$ in units of the inverse coupling $t^{-1}$.} \label{fig:edgeChain}
\end{figure}

\begin{figure}[t]
\includegraphics[clip,width=\linewidth]{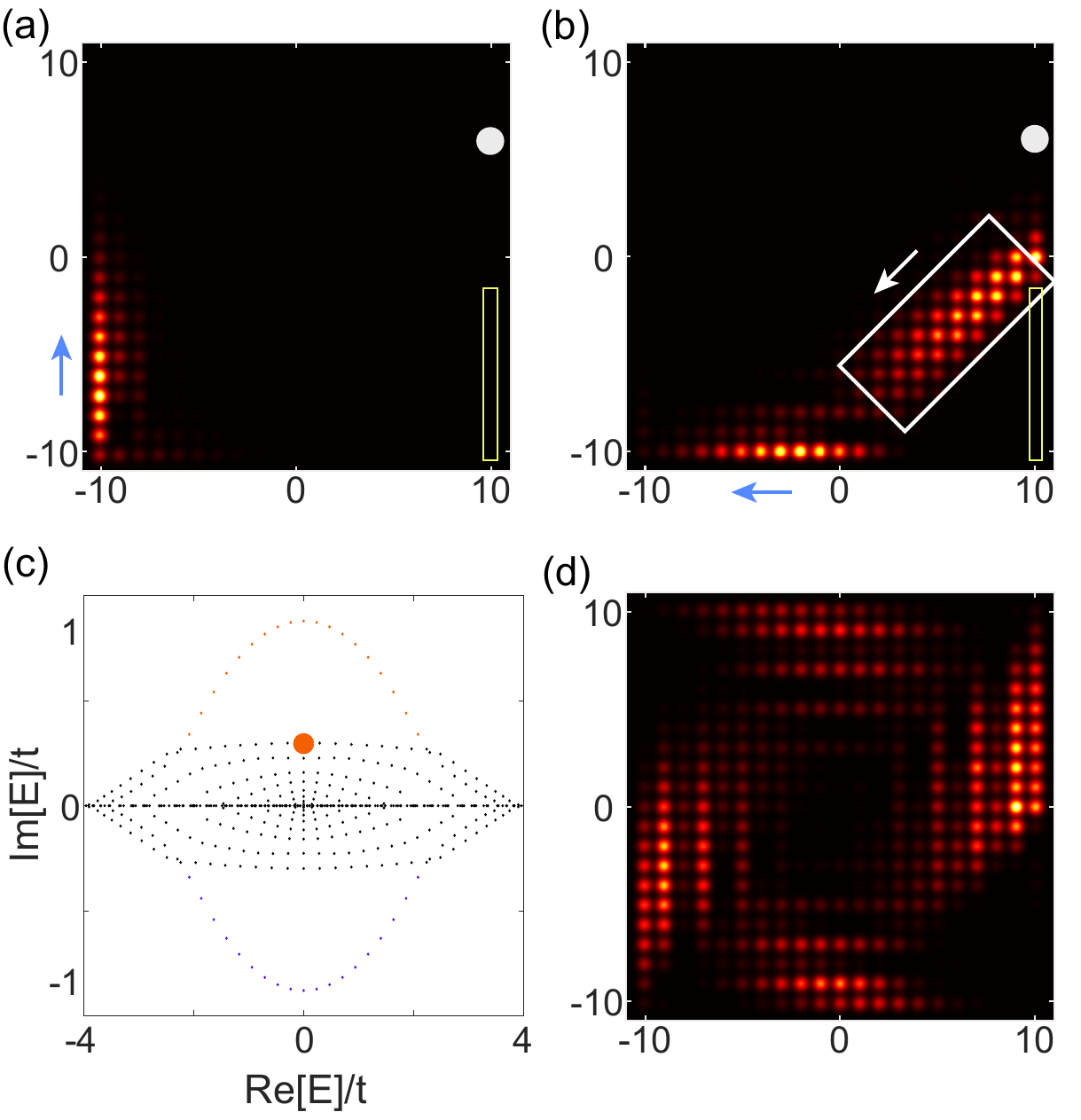}
\caption{(a,b) Same as Fig.~\ref{fig:edgeChain} but with the bulk present. The pulse at $\tau=4.4 t^{-1}$ is almost identical to Fig.~\ref{fig:square_dynamics}(b). (c) The complex energy spectrum with the topological defect and $s=2$. The wave function of the high-gain bulk-like mode marked by the large orange dot is shown in (d).} \label{fig:topoDefect}
\end{figure}

Not only can bulk excitation initialize a CEMIC but also assist a CEMIC bridging sizable ``topological'' defects in our system. As an example, we replace the asymmetric couplings by symmetric couplings in the bottom half of the right edge (see the yellow ribbon in Fig.~\ref{fig:edgeChain}). In the absence of the bulk, i.e., in a 1D chain around the perimeter of the square domain, this defect acts similarly to an OBC and leads to a non-Hermitian skin-like effect, i.e., all eigenstates are localized at the middle of the right edge, with a long tail extending upward and around the corner (due to the positive and small $s=2$) and a very short tail extruding downward (due to the topological defect). This topological defect hinders the propagation of the wave packet, which can be seen by comparing Figs.~\ref{fig:square_dynamics}(c) and \ref{fig:edgeChain}(b) taken at the same time. The wave packet then ``tunnels'' to the bottom of the right edge and continues its circulation [see Fig.~\ref{fig:edgeChain}(a)]. In the presence of the bulk, however, the same wave packet is able to cut the corner, traverse to the middle of the bottom edge [see Fig.~\ref{fig:topoDefect}(b)], and resume its edge propagation. In this case, the complex energy spectrum is just slightly perturbed from the original one without the topological defect [cf. Figs.~\ref{fig:square}(c) and \ref{fig:topoDefect}(c)], and this corner-cutting behavior is facilitated by the high-gain bulk-like modes, such as the one shown in Fig.~\ref{fig:topoDefect}(d).    

\section{Conclusion and Discussion}
In summary, we showed that topological chiral edge bands can exist in the continuum of a topologically trivial bulk, contrary to the Hermitian case. The topological origin of these CEMICs is inherited from the Hatano-Nelson model with PBC along the edge, which manifests in the strong asymmetric coupling limit. Nevertheless, the edge localization of CEMICs is not due to the non-Hermitian skin effect typically associated with the Hatano-Nelson model with OBC. Instead, it can be attributed to a PT transition without explicit gain and loss. Their onset requires a critical value of coupling asymmetry, which is given by the golden ratio between the edge couplings and their geometric average. 

The effective gain obtained at these asymmetric coupling junctions leads to a strong amplification as a CEMIC wave packet circulates. To avoid gain saturation and material breakdown, one could use uniform material loss to balance or reduce this strong amplification. Note that this compensation does not affect the robustness of CEMICs, because this additional loss applies to all modes in our system. In other words, it does not close the imaginary band gap or open a real band gap. Therefore, scattering caused by defects into the opposite circulation direction and the bulk is still strongly suppressed, and bulk modes are still able to assist CEMICs traversing topological defects along the edge. Our finding heralds new domains where topological states are unexpected in the Hermitian tenet, which can be conveniently explored in photonics, acoustics, electronics, mechanics, and related systems.

\begin{acknowledgments}
This work is supported by the National Science Foundation (NSF) under Grant No. ECCS-2602409 and a PSC-CUNY research award from the City University of New York under Grant No. GR-00018039. 
\end{acknowledgments}

\appendix

\section{Strong asymmetric couplings along the edges}
\label{app:couplings}

\begin{figure}[t]
\includegraphics[clip,width=\linewidth]{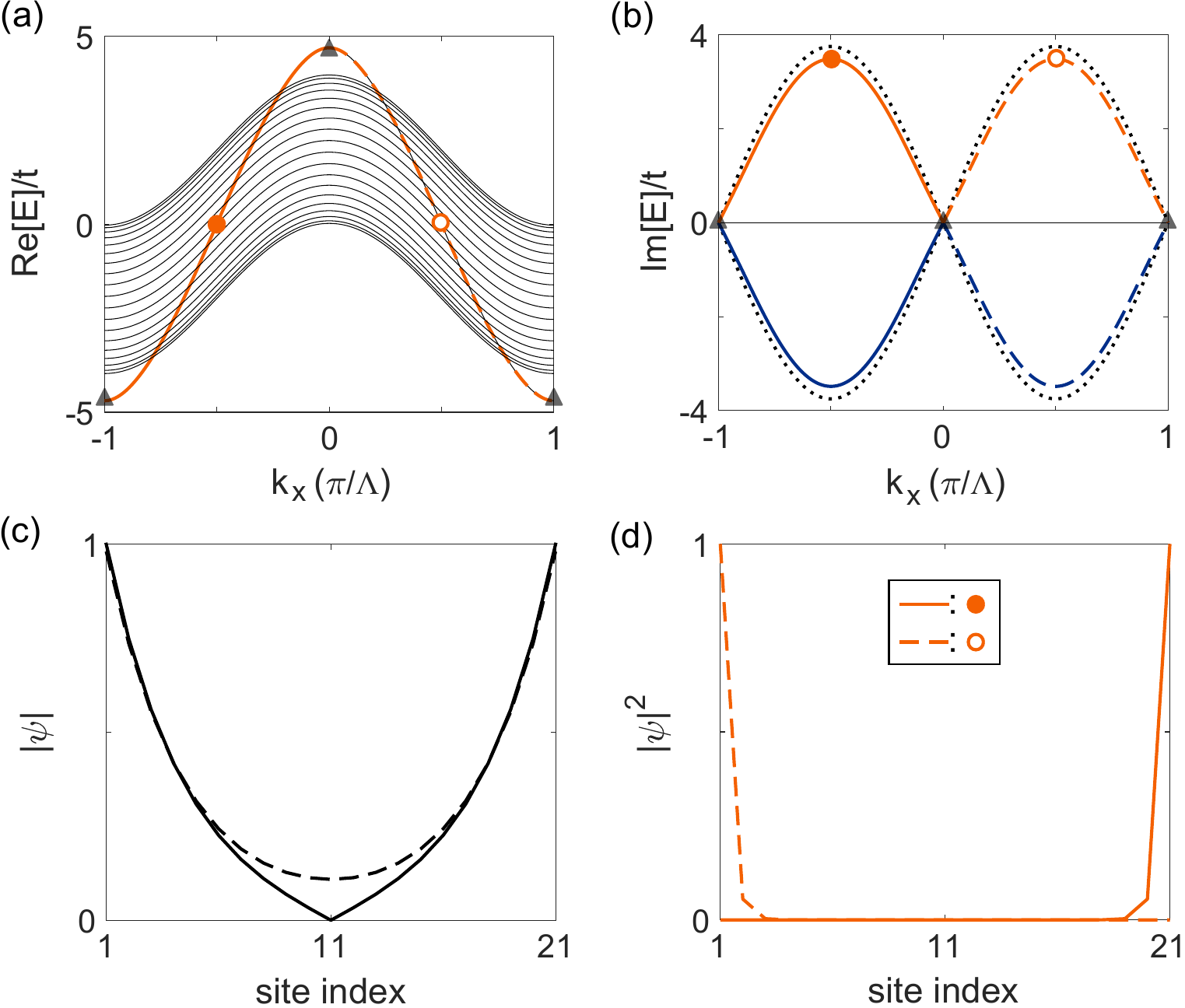}
\caption{\textbf{CEMICs across almost the entire Brillouin zone in a ribbon}. (a,b) Same as Figs.~\ref{fig:1}(e) and \ref{fig:1}(d) but with $s=4$. Triangles now show almost degenerate Hermitian edge modes. The two wave functions at each triangle are indistinguishable by eye at $s=4$, and hence we plot them at a lower $s=3$ instead at $k_x=-\pi/\Lambda$ in (c). Those of the two CEMICs marked by the large dots in (a,b) are shown in (d).} \label{fig:2}
\end{figure}

As we mentioned in the main text, if we further increase $s$ in our ribbon model, the PT-broken phases broaden and push the EPs outward toward $k_x=0$ and $\pm\pi/\Lambda$. Note that exactly at these high-symmetry points, the effective gain and loss $\gamma$ given by Eq.~(\ref{eq:ImE}) of the main text vanishes, and the ribbon Hamiltonian becomes Hermitian:
\be
H_0 = \pm t\begin{pmatrix}
(s+1/s) & 1 & & & \\
         1 & 2 & 1 & & \\
           & \ddots & \ddots & \ddots & \\
           & & 1 & 2 & 1 \\
           & &  & 1  & (s+1/s)
\end{pmatrix}.\label{eq:H0}
\ee 
For $s\gg1$, the two ends of the ribbon are effectively detached from the bulk, this time not because of the imaginary band gap but a real one. Furthermore, the two edges are weakly coupled via the bulk, forming a pair of symmetric and anti-symmetric edge modes [see Fig.~\ref{fig:2}(c)]. Their coupling is so weak and they appear as degenerate [triangles in Figs.~\ref{fig:2}(a) and \ref{fig:2}(b)], but they still have a finite real energy separation, which is on the order of $10^{-7}t$ with $s=4$ and $N_y=21$.    

The strong edge localization displayed by the CEMICs in this case [see Fig.~\ref{fig:2}(d)] also makes the effective gain and loss $\gamma$ in Eq.~(\ref{eq:ImE}) of the main text a much better bound in the entire Brillouin zone [see Fig.~\ref{fig:2}(b)]. This bound was derived using the 1D Hatano-Nelson model with PBC, which also gives
\be 
\re{E_b} = (s+1/s)t \sin k_x\Lambda
\ee    
and the group velocity
\be 
\frac{d E_b}{dk} = (s+1/s) \Lambda t \cos k_x\Lambda.
\ee 
At the place of maximum gain in the Brillouin zone for the CW CEMIC, i.e., $k_x=-\pi/2\Lambda$, the group velocity becomes $(s+1/s) \Lambda t$.

\section{EP of the ribbon model at $k_x=\pm \pi/2\Lambda$}
\label{app:EP}

The ribbon model at $k_x=\pm \pi/2\Lambda$ given by Eq.~(\ref{eq:Hc}) has PT symmetry as well as non-Hermitian particle-hole (NHPH) symmetry \cite{zeromodeLaser,qi_defect_2018}. While the former leads to $E_\mu=E_\nu^*$ for a pair of energy eigenvalues (where $\mu,\nu$ can be equal), the latter leads to $E_\mu=-E_{\mu'}^*$ (where $\mu,\mu'$ can be equal). In other words, the spectrum of this model is symmetric about both the real and imaginary axes of the complex energy plane, which also implies a non-Hermitian chiral symmetry \cite{rivero_chiral_2021} $E_\nu = -E_{\mu'}$. In our case here, we find that the EP first appears at the origin, i.e., with $E=0$ at the critical value $s_c$ where the PT transition that leads to CEMICs take place.

To find this critical value $s_c$, we note that with $E=0$, the eigenvalue problem of the ribbon Hamiltonian $H_c$ [given by Eq.~(\ref{eq:Hc}) for $k_x=-\pi/2\Lambda$] reduces to a linear homogeneous recurrence relation
\be
0 = \psi_{n+1} + \psi_{n-1}\quad(n=2,3,\ldots,N_y-1) \label{eq:recur}
\ee
with the following OBCs along the $y$-direction:
\begin{align}
0 & =  i(s-1/s)\psi_{1} + \psi_2,\\
0 & = - i(s-1/s)\psi_{N_y} + \psi_{(N_y-1)}.
\end{align}
The recurrence relation (\ref{eq:recur}) can be solved using the ansatz 
\be
\psi_n = \beta_+ b_+^{n-1} + \beta_- b_-^{n-1} \quad(n=1,2,\ldots,N_y)\label{eq:psi_recur}
\ee
where $b_\pm$ are the solutions of 
\be
b^2 + 1 = 0,
\ee
i.e., the characteristic polynomial of Eq.~(\ref{eq:recur}). Simply, we find $b_\pm = \pm i$. 

The two boundary conditions mentioned above then lead to
\begin{align}
0 & = i\gamma(\beta_+ + \beta_-) + it(\beta_+ - \beta_-),\label{eq:recur_bd1} \\
0 & = -i\gamma[\beta_+ i^{N_y-1} + \beta_- (-i)^{N_y-1}] \nonumber \\
& \hspace{2cm} + t[\beta_+ i^{N_y-2} + \beta_- (-i)^{N_y-2}].
\end{align}
For $\beta_\pm$ to have a non-zero solution, the determinant of 
\be
M\equiv
\begin{pmatrix}
\gamma+t & \gamma-t \\
(\gamma+t)i^{N_y-1} & (-)^{N_y-1}(\gamma-t)i^{N_y-1} 
\end{pmatrix}
\ee
must vanish, i.e.,
\be
(-)^{N_y-1}(\gamma^2-t^2) = \gamma^2-t^2.\label{eq:EP_cond1}
\ee
Since $\gamma$ appears in its squared form, this relation applies to the ribbon Hamiltonian $H_c$ at both $k_x=\mp\pi/2\Lambda$.

If $N_y$ is even, this condition is satisfied when $\gamma=t$ (again $\gamma,t>0$ by definition), or equivalently,
\be
s - 1/s = 1.\label{eq:s_golden}
\ee   
The positive solution of $s$ is then given by the golden ratio
\be
s = \frac{1+\sqrt{5}}{2}.
\ee

If $N_y$ is odd, then the EP condition (\ref{eq:EP_cond1}) is always satisfied, which is not surprising given the non-Hermitian chiral symmetry mentioned above and the Limb theorem \cite{rivero_chiral_2021}, i.e., there is always one zero mode (satisfying $E=0$) when the lattice size is odd. Therefore, we need another EP condition to identify the critical value $s_c$.

To this end, we first note that the ribbon Hamiltonian $H_c$ is symmetric, at both $k = \pm\pi/2\Lambda$ . This is because the off-diagonal elements in it describe the \textit{symmetric} couplings in the vertical direction; the asymmetric couplings in the horizontal direction only appear through PBC in the same direction, i.e., as diagonal elements in the ribbon Hamiltonian. For this (and other) symmetric non-Hermitian Hamiltonian, the coalesced wave function at its EP is self-orthogonal in the unconjugated inner product \cite{rivero_pseudochirality_2020}. Here the general form of the wave function for a zero mode, i.e., Eq.~(\ref{eq:psi_recur}), can be written as
\be
\Psi = [(\beta_+ + \beta_-), i(\beta_+ - \beta_-), -(\beta_+ + \beta_-), -i(\beta_+ - \beta_-), \dots]^T,\nonumber
\ee
which repeats every four lattice sites. When $N_y$ is even, its vanished unconjugated inner product with itself gives
\be
(\beta_+ + \beta_-)^2 - (\beta_+ - \beta_-)^2 = 0,
\ee
leading to 
\be
\beta_+\beta_-=0.
\ee
Plug it back into the boundary condition (\ref{eq:recur_bd1}) and we again find $\gamma=t$ and Eq.~(\ref{eq:s_golden}). When $N_y$ is odd though, the vanished unconjugated inner product of the wave function above leads to
\be
(N_y+1)(\beta_+ + \beta_-)^2 - (N_y-1)(\beta_+ - \beta_-)^2 = 0,
\ee 
and together with the boundary condition (\ref{eq:recur_bd1}), we find
\be
\gamma = t\sqrt{\frac{N_y+1}{N_y-1}},
\ee
which gives again $\gamma=t$ and Eq.~(\ref{eq:s_golden}) when $N_y\gg1$. By also noticing that $\gamma=t$ leads to $\beta_+=0$, the wave function at the EP is then given by
\be
\Psi = [1, -i, -1, i, \dots]^T,
\ee
which has a uniform probability or intensity distribution along the ribbon.

\bibliography{references}

\end{document}